\documentclass[12pt]{aastex631}
\usepackage{multirow}
\usepackage{color}
\usepackage{amsmath}
\usepackage[percent]{overpic}
\usepackage{rotating}
\usepackage{subfigure}
\usepackage{makecell}
\definecolor{DarkGreen}{rgb}{0.0,0.4,0.0}  
\newcommand{\B}[1]{{\color{blue}{#1}}}
\newcommand{\R}[1]{{\color{red}{#1}}}
\newcommand{\G}[1]{{\color{DarkGreen}{#1}}}
\newcommand{\M}[1]{{\color{magenta}{#1}}}
\newcommand{\C}[1]{{\color{cyan}{#1}}}

\usepackage[normalem]{ulem}
\usepackage{soul}
\usepackage{cancel}

\usepackage[titletoc]{appendix} 
\makeatletter
\AtBeginDocument{%
  \@ifundefined{MakeLineNo}{}{

  }
}
\makeatother
\shorttitle{Size Distributions of Arcsecond-Scale Ribbon Properties}
\shortauthors{Zhang et al.}
\graphicspath{{./}{figures/}}

\begin{document}

\title{Size Distributions of Arcsecond-Scale Properties of Solar Flare Ribbons}

\author[0009-0001-1500-5053]{Yue Zhang}
\affiliation{CAS Key Laboratory of Geospace Environment, Department of Geophysics and Planetary Sciences, University of Science and Technology of China, Hefei 230026,
People’s Republic of China}

\author[0000-0003-4618-4979]{Rui Liu}
\affiliation{CAS Key Laboratory of Geospace Environment, Department of Geophysics and Planetary Sciences, University of Science and Technology of China, Hefei 230026,
People’s Republic of China}
\affiliation{Mengcheng National Geophysical Observatory\\
University of Science and Technology of China, Hefei, 230026, China}

\author[0000-0002-9865-5245]{Wensi Wang}
\affiliation{CAS Key Laboratory of Geospace Environment, Department of Geophysics and Planetary Sciences, University of Science and Technology of China, Hefei 230026,
People’s Republic of China}
\affiliation{Mengcheng National Geophysical Observatory\\
University of Science and Technology of China, Hefei, 230026, China}

\author[0009-0006-7051-0438]{Junyan Liu}
\affiliation{CAS Key Laboratory of Geospace Environment, Department of Geophysics and Planetary Sciences, University of Science and Technology of China, Hefei 230026,
People’s Republic of China}

\correspondingauthor{Rui Liu}
\email{rliu@ustc.edu.cn}

\begin{abstract}

Solar flare ribbons are believed to map the footpoints of newly reconnected magnetic flux tubes, therefore shedding light on the reconnecting current sheet, which is rarely observed by direct imaging or spectroscopy. Here we study the detailed evolution of flare ribbons down to the arcsecond scale for 10 flares characterized by the classic double ribbons. Identifying the flaring pixels by combining the intensity variances of the UV filter ratio and intensity threshold, we found that the waiting time distributions of the flaring pixels are well described by power laws, distinct from those in the preflare or quiet-Sun regions, and that the power-law slopes are generally consistent with those predicted by nonstationary Poisson processes in the nonlinear regime or by the 2D/3D self-organized criticality (SOC) model. The size distributions for flaring duration also follow power laws but the slopes are more scattered. In contrast, the size distributions for other parameters, including peak intensity, energy, and radial magnetic field strength of the flaring pixels, deviate from power laws, and the estimated slopes significantly differ from the SOC predictions. These results suggest that a nonstationary Poisson process or an avalanche-like process might be ongoing in the temporal dimension in the reconnecting current sheet, but in other aspects, e.g., space- and energy-wise, the avalanche is likely modulated by other physical processes or the fine structures of the reconnecting current sheet.

\end{abstract}

\section{Introduction} \label{S-intro}
Solar flares are observed as a sudden, local brightening over the entire electromagnetic spectrum in the solar atmosphere \citep{fletcherObservationalOverviewSolar2011,benzFlareObservations2017}. In the standard flare model \cite[]{carmichaelProcessFlares1964, sturrockModelSolarFlares1968, hirayamaTheoreticalModelFlares1974, koppMagneticReconnectionCorona1976}, reconnection occurs at a very thin current sheet, generated by the inflow of oppositely directed magnetic field lines underneath a rising magnetic flux rope. The reconnection energizes a large number of electrons, which propagate along newly reconnected magnetic flux tubes to the chromosphere, where they are thermalized and manifested as flare ribbons. Thus, flare ribbons map the footpoints of magnetic flux tubes reconnecting at the flaring current sheet in the corona. Since it is difficult to observe the magnetic reconnection directly, flare ribbons can be a powerful tool to diagnose the reconnection process in the corona and the internal structure of the current sheet \cite[e.g.,][]{wangStudyRibbonSeparation2003,nishizukaPOWERLAWDISTRIBUTIONFLARE2009,janvierEvolutionFlareRibbons2016,Wang2017,wyperFlareRibbonFine2021,Gou2023}.

Various physical parameters of solar flares, such as peak intensity, flare duration, fluence, waiting time between discrete events, are often found to follow power-law distributions, which may have important implications for the physical processes and physical scaling laws of flares \citep{boffettaPowerLawsSolar1999, wheatlandOriginSolarFlare2000, nishizukaPOWERLAWDISTRIBUTIONFLARE2009, aschwandenRECONCILIATIONWAITINGTIME2010, aschwanden25YearsSelfOrganized2016, leiSolarFlaresOriginating2020, aschwandenCorrelationSunspotNumber2021,Yang&Liu2025}.  
In particular, Waiting-time distributions (WTDs) of solar flares may help distinguish different types of random processes. \cite{aschwandenCorrelationSunspotNumber2021} compiled a comprehensive list of published WTDs of solar flares, CMEs, and radio bursts. Most of the WTDs display a power law distribution function $N(\Delta t)\,d\Delta t\propto \Delta t^{-\alpha}\,d\Delta t$ with $\alpha$ in the range of [0.75, 3]. \cite{nishizukaPOWERLAWDISTRIBUTIONFLARE2009} studied the \ion{C}{4} kernels on the flare ribbons in UV 1600~{\AA} images in a single flare event and found that not just the peak intensity and duration of flare kernels but also their waiting times follow power-law distributions. Such a power law decay over a wide range of scales indicates scale invariance. The scale invariance is a well-known hallmark of complex systems, which can be explained by either self-organized criticality \cite[SOC;][]{aschwandenRECONCILIATIONWAITINGTIME2010, aschwandenAUTOMATEDSOLARFLARE2012, aschwandenMACROSCOPICDESCRIPTIONGENERALIZED2014}), non-stationary Poisson process \citep{wheatlandOriginSolarFlare2000, wheatlandCoronalMassEjection2003}, or MHD turbulence \citep{boffettaPowerLawsSolar1999, lepretiSolarFlareWaiting2001, grigoliniDiffusionEntropyWaiting2002}. \cite{wheatlandCoronalMassEjection2003} demonstrated that a non-stationary Poisson process with an exponentially growing or decaying flare rate, $\lambda(t)$, generates a WTD with a power-law tail $\sim \Delta t^{-3}$.
\cite{aschwandenRECONCILIATIONWAITINGTIME2010} used $\delta$-functions as the flare rate $\lambda(t)$, which yields a WTD with a power-law tail $\sim \Delta t^{-2}$.
With a generalized distribution of event rate, $f(\lambda) = A\lambda^{-\alpha}\exp(-\beta\lambda)$, \cite{Li2014} derived a WTD with a power-law tail $\sim\Delta t^{-(3-\alpha)}$, with $ 0\le\alpha<2$. Alternatively, \cite{Aschwanden2021poisson} parameterized the event rate function with an exponent $p$ so that $\lambda(t)\propto t^p$, which yields a power-law slope $\alpha=2+1/p$ in the range of $[2, 2.5]$ in the nonlinear regime ($p\ge2$). \cite{boffettaPowerLawsSolar1999} proposed that flares may represent intermittent dissipative events of MHD turbulence;  \cite{lepretiSolarFlareWaiting2001} argued that the observed WTD, which can be reproduced by a L\'{e}vy function, may suggest that the turbulence associated with solar flares is statistically self-similar in time and hence possesses some memory.

SOC was first proposed by \cite{bakSelforganizedCriticalityExplanation1987} to unify the spatial and temporal fractals that are ubiquitous in nature and characterized by a power-law (frequency) size distribution, such as the $1/f$ power spectra. \cite{luAvalanchesDistributionSolar1991} introduced SOC to the study of solar flares, piloting the application of the SOC theory in solar physics and astrophysics. Generally, \emph{``SOC theory describes a critical state of a nonlinear energy dissipation system that is slowly and continuously driven towards the critical value of a system-wide instability threshold, producing scale-free, fractal-diffusive, and intermittent avalanches with power-law size distributions; the driving process towards the critical state is absent of any external parameter adjustment and characterized by scale invariance, i.e., the power-law size distribution of avalanches is the same in different system sizes''}\citep[p55]{aschwandenMACROSCOPICDESCRIPTIONGENERALIZED2014}. 
More recently, \cite{aschwandenReconcilingPowerlawSlopes2022} found that the standard SOC model is able to predict the power-law slopes of the size distribution of solar flare energies down to the nanoflare regime. \cite{Aschwanden2022ApJ} further synthesized data including various parameters of solar and stellar flares from 162 publications and found that the majority of them are consistent with the standard SOC model. These results suggest that the SOC theory is promising for deepening our understanding of solar flares.

In this paper, we study the detailed evolution of flare ribbons down to individual pixels in 10 selected events from 2011 to 2023. We do not attempt to identify the flare kernels, but instead treat each individual pixels on the flare ribbons as the unresolved footpoints of newly reconnected magnetic flux tubes, whose sizes could be as small as 80--200 km as revealed by high-resolution H$\alpha$ observations \citep{Jing2016}, smaller than the pixel size ($0''.6$ or 430 km) of the UV images used in this study. In the remainder of the paper, we accurately identify the flare ribbons with an improved threshold method (\S\ref{sec:methods}). We then investigate the power-law indices of the size distributions for various physical parameters derived from the light curve of each individual pixel on the flare ribbons, and compare the obtained indices with those anticipated by the standard SOC model (\S\ref{sec:results}), which is an analytical physical model derived from first principles (see Appendix~\ref{app-theory}). We make concluding remarks in \S\ref{sec:conclusion}.  

\section{Methods} \label{sec:methods}

\subsection{Instruments}
The selected events are all classic two-ribbon flares occurring within 45 deg from the disk center (Table~\ref{T-list_of_events}). To study the evolution of flare ribbons, we used UV 1600 and 1700 \AA\ images with a pixel scale of $0''.6$ and a temporal cadence of 24 seconds, taken by the Atmospheric Imaging Assembly \cite[AIA;][]{lemenAtmosphericImagingAssembly2012} on board the Solar Dynamics Observatory \cite[SDO;][]{pesnellSolarDynamicsObservatory2012}. We also used the Helioseismic and Magnetic Imager \cite[HMI;][]{scherrerHelioseismicMagneticImager2012, schouDesignGroundCalibration2012}, particularly the hmi.sharp\_cea\_720s data series of the Spaceweather HMI Active Region Patch (SHARP), to study the magnetic field at flare ribbons. The X1.6-class flare on 2014 September 10 (Event \#5) was also observed by the Interface Region Imaging Spectrograph \cite[IRIS;][]{depontieuInterfaceRegionImaging2014} Slit-Jaw Imager (SJI) at 1400~{\AA}.  Since IRIS images are superior in temporal cadence (19 sec) and pixel scale ($0''.167$) to AIA, we compared the results from the two instruments to check potential effects arising from different image resolution. 

\subsection{Data Reduction}  
\subsubsection{Identifying flaring pixels}
\begin{figure} 
\centerline{\hspace*{0.015\textwidth}
\hspace*{-0.03\textwidth}
\includegraphics[width=1.08\textwidth,clip=]{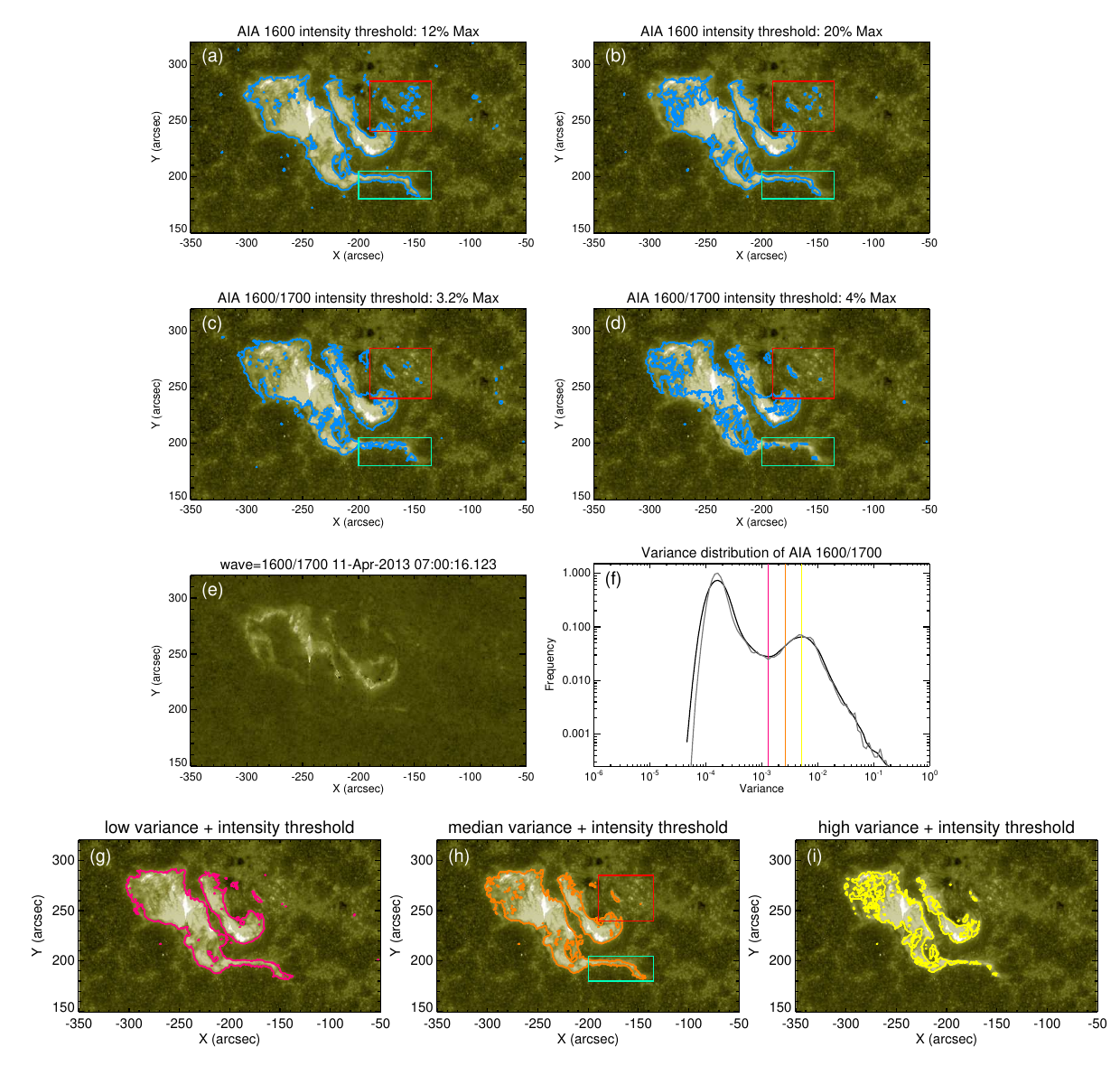}}
\caption{Identification of flare ribbons in SDO/AIA's UV passbands for the 2013 April 11 event. The background image in panels (a--d) and (g--i) is a 1600~{\AA} synoptic map of flare ribbons, in which each pixel is shown by its maximum intensity during the flare period. The superimposed contours indicate the extracted flare ribbons. Flare ribbons are identified by a conventional threshold method with \B{2 different thresholds applied to 1600~{\AA} images (a \& b) and to 1600/1700~{\AA} ratio images (c \& d), respectively.} The threshold is chosen by a fraction of the maximum intensity of all pixels in the presented field of view (FOV) during the flare period, and those pixels with peak intensity above the threshold are taken as part of the flare ribbons. \B{The red box marks an off-ribbon region with relatively intense plage brightening, and the green box a segment of the flare ribbon.} (e) A map of AIA 1600/1700~{\AA} intensity ratio. (f) Size distribution of intensity variances for all the pixels in the FOV of Panel (e) during the flare period; the original data before being smoothed is shown by the gray line. Vertical lines correspond to three different variance thresholds set in (g--i), namely, the local minimum between the two humps (pink), the peak of the 2nd hump (yellow), and their median (orange). \B{An intensity threshold is further applied to extract the flare ribbons (see the text).} 
\label{F-ribbon_contour} }  
\end{figure}

The identification of flare ribbons is often plagued by the chromospheric networks and especially the plages, which are relatively bright and host many randomly brightening points. First, we take the ratio of AIA 1600 and 1700~{\AA} images to remove chromospheric networks and plages \cite[Fig.~\ref{F-ribbon_contour}e;][]{dudikSLIPPINGMAGNETICRECONNECTION2016, lorincikManifestationsThreedimensionalMagnetic2019}. Next, we calculate the intensity variance of the light curve of the 1600/1700 ratio for each pixel and obtain the frequency distribution of these variances. The distribution as shown in Fig.~\ref{F-ribbon_contour}f is further smoothed by the Savitzky-Golay convolution method with IDL procedures, \texttt{SAVGOL} and \texttt{CONVOL}, and two humps can be clearly seen: the 1st hump corresponding to smaller variances is attributed to the background, while the 2nd hump corresponding to larger variances is attributed to flare ribbons. Through trial and error, we found that those pixels with variances larger than a threshold, which ranges from the local minimum between the two humps to the local maximum of the 2nd hump, compare favorably with the visually recognized flare ribbons in the \emph{synoptic map}, in which each pixel is shown by its maximum intensity during the flaring period to highlight the ribbon-swept area. We hence take the median of the above-mentioned variance range as the threshold value to identify flare ribbons and take the lower and upper bound of the threshold range to estimate the uncertainties of the identification (Fig.~\ref{F-ribbon_contour}(g--i)). However, some pixels off the ribbons may have a large variance in the 1600/1700 ratio but a low intensity in the 1600 or 1700~{\AA} passband. We further apply an intensity threshold to filter out such undesirable pixels. We take the average intensity of each individual pixel over a 30-min interval immediately before the flare onset as its own preflare background $P_0$, and require that a flaring pixel must have a peak intensity $P$ five times larger than $P_0$.

As a comparison, we also used the conventional threshold method (Fig.~\ref{F-ribbon_contour}(a--d)) to extract flare ribbons using 1600~{\AA} images \B{(Fig.~\ref{F-ribbon_contour}(a \& b))} and the 1600/1700 filter ratio \B{(Fig.~\ref{F-ribbon_contour}(c \& d))} at two different thresholds. At the lower threshold (Fig.~\ref{F-ribbon_contour}(a \& c)), some plage brightenings are misidentified as flare ribbons; at the higher threshold (Fig.~\ref{F-ribbon_contour}(b \& d)), less plage brightenings but also less ribbon brightenings are included in the extracted flare ribbons. \B{For a detailed comparison between the different methods, we marked two representative regions, an off-ribbon region with relatively intense plage brightening (red box) and a segment of the flare ribbon (green box). By carefully selecting an intensity threshold for 1600~{\AA} images, one can successfully extract the flare ribbons, but inevitably picking up numerous pixels belonging to plages (Fig.~\ref{F-ribbon_contour}a). A higher intensity threshold may pick up less plage pixels, but miss more ribbon pixels (Fig.~\ref{F-ribbon_contour}b). Applying an intensity threshold to 1600/1700 filter ratio can generally suppress the plage brightening, but it performs poorly on the flare ribbon (Fig.~\ref{F-ribbon_contour}(c \& d)). Our procedure combining the variance and intensity threshold (Fig.~\ref{F-ribbon_contour}h) strikes a better balance by identifying flare ribbons accurately, while significantly reducing the background noise introduced by plages.}

\subsubsection{Deriving the waiting time distribution of flaring pixels}
\begin{figure} 
\centerline{\hspace*{0.015\textwidth}
\hspace*{-0.03\textwidth}
\includegraphics[width=1.0\textwidth,clip=]{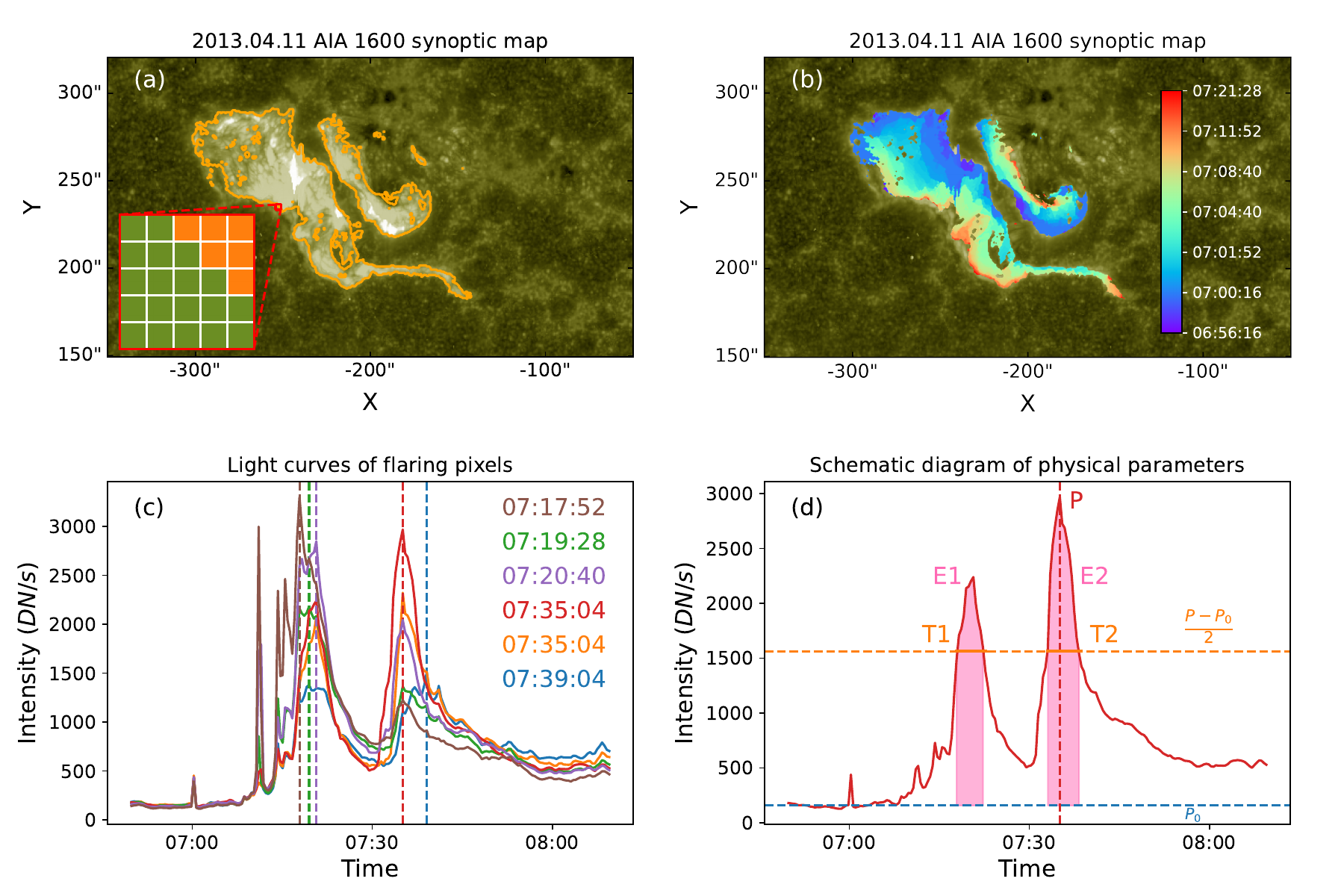}}
\caption{Schematic diagram to illustrate the derivation of physical parameters from light curves of flaring pixels. Panel (a) shows the synoptic map of flare ribbons on 2013 April 11. The red square marks a representative grid cell of $5\times5$ pixels. The inset zooms into this cell to show 6 pixels on the flare ribbon, as indicated in orange colors. Panel (b) shows a map of color-coded peak times of the light curves of flaring pixels. Panel (c) shows the light curves of the 6 flaring pixels in the representative grid cell in (a), with peak times annotated and marked by vertical dashed lines. Panel (d) shows how peak intensity ($P$), duration ($T=T_1+T_2$), and energy ($E= E_1+E_2$) are derived from a sample light curve. The preflare background $P_0$ is defined as the average over 30 minutes before the flare onset. The duration $T$ is determined as the full width at $(P-P_0)/2$.} 
\label{F-illusion}  
\end{figure}

The obtain the WTD of flaring pixels in a single flare event, we adopted the approach of \cite{nishizukaPOWERLAWDISTRIBUTIONFLARE2009}: we divide the flare region by a uniform grid with each cell of the size $5\times5$ pixels. Within each cell, we obtain light curves of the pixels that are identified to be on the flare ribbons and take the consecutive time intervals between the peak times of these light curves as waiting times; in other words, we assume that there is no long-range spatio-temporal correlation between flaring pixels that belong to different grid cells, which is often the case for many complex systems \cite[]{Sanchez&Newman2018}. Eventually we collect all the waiting times from each individual grid cells and `lump' them together to obtain \B{the overall WTD; i.e., the probability of finding the waiting time $\Delta t$ in the  frequency distribution of waiting times for the whole flare region is given by
\begin{equation}
    P(\Delta t)\,dt = \frac{1}{N} \sum\limits_{i=1}^N  f_i(\Delta t)\,dt, 
\end{equation}
in which $f_i$ gives how often the waiting time $\Delta t$ appears in the $i$th cell of the total number of $N$ cells.}    
As an example, Figure~\ref{F-illusion}a shows a cell (red rectangle) taken from the $5\times5$ grid. 6 of the 25 pixels are identified as flaring pixels (orange) on the flare ribbons (inset of Figure~\ref{F-illusion}a). The waiting times counted in this cell are the four consecutive time intervals between the peak times of the six flaring pixels (Figure~\ref{F-illusion}c), two of which has the same peak time.

We also tested on different grid patterns ($7\times7$ and $9\times9$ pixels), and the resultant power-law slopes of the WTDs are shown in Table~\ref{T-grid}, with their different ranges indicated as colored stripes in Fig.~\ref{F-slope_vs_theory}. One can see that with larger cells, the WTD typically becomes slightly steeper, because more flaring pixels are taken to be connected to each other. If a grid cell is too small, a flaring pixel would have too few neighbors; if a grid cell is too large, it includes pixels located far away from each other, which does not only reduce the number of possible waiting times because of the coarse temporal cadence (24~s), but also violates the assumption as mentioned above. In the extreme case of no grid at all, the waiting times would be lumped into only a few bins (cf. Figure~\ref{F-illusion}b). Hence, to strike a balance, the uniform grid with the cell size of 5 by 5 pixels, which yields a sufficient number of waiting times for statistics, is employed in this study.

\subsubsection{Deriving the size distributions of other parameters}
Without partitioning the images by grids, we investigated other properties of flare ribbons. For each flaring pixel on the ribbons, we obtained the peak intensity $P$ in units of DN~s$^{-1}$, the flare duration $T$ of the light curve in units of second, the area below the light curve above the preflare background in units of DN as a proxy of the energy $E$, which is a product of the mean flux $F$ and the time duration $T$ in the standard SOC model \cite[]{Aschwanden2022ApJ}, and the radial magnetic field strength $B_r$ in units of Gauss. To obtain $B_r$ at flare ribbons, the synoptic map of AIA 1600~{\AA} images is performed the Lambert Cylindrical Equal-Area (CEA) projection and interpolation to match a vector magnetogram of the SHARP data immediately before the flare onset. In one representative light curve (Figure~\ref{F-illusion}d), we mark the peak intensity ($P$) and two time intervals ($T_1$ and $T_2$) above the $(P-P_0)/2$ level. Since $P>5P_0$ is required for the identification of flaring pixels, the uncertainty introduced by $P_0$ is less than 20\%. For $T_1$ and $T_2$, we calculate the area $E_1$ and $E_2$, respectively, which are below the light curve but above the preflare background $P_0$; the sum of $E_1$ and $E_2$ is taken a proxy of the energy $E$, and the sum of $T_1$ and $T_2$ is taken as the flare duration $T$. 

\subsubsection{Power-law fitting}
The frequency-size distributions of the parameters as defined above are then fit with a power-law function with the maximum likelihood estimation (MLE) method \B{(\texttt{powerlaw.py} in Python)}, which is accurate, robust, and unbiased over other graphical methods based on logarithmic scale linear fittings \citep{goldsteinProblemsFittingPowerlaw2004, clausetPowerLawDistributionsEmpirical2009}. \B{For a continuous random variable $x$ bounded by $x_{\min}$, the probability density function (pdf) $p(x)$ taking the form of power law is given as follows, 
\begin{equation}
    p(x;\alpha,x_{\min})=Cx^{-\alpha}=\frac{\alpha-1}{x_{\min}} \left(\frac{x}{x_{\min}}\right)^{-\alpha}, \label{eq:powerlaw}
\end{equation}
where $\alpha$ is the power-law index and $C$ is the normalization constant. The MLE estimation of $\alpha$ gives \begin{equation}
    \hat{\alpha} = 1+n\left[\sum_{i=1}^{n}\ln{\frac{x_i}{x_{\min}}}\right]^{-1}.    
\end{equation}
However, when $x$ is discretely sampled, that the pdf must satisfy $\sum_{x=x_{\min}}^{\infty} Cx^{-\alpha}=1$ gives
\begin{equation}
    p(x;\alpha,x_{\min})= \frac{x^{-\alpha}}{\zeta(\alpha, x_{min})},
\end{equation}
where $\zeta(\alpha, x_{\min})= \sum_{n=0}^\infty(n+x_{\min})^{-\alpha}$ is the generalized zeta function \cite[]{clausetPowerLawDistributionsEmpirical2009}. 
In our work, $\Delta t$ and $T$ are discrete since the image cadence is 24~s. But other parameters, $P$, $E$, and $B$ are  continuous.} The fitting interval is generally determined by minimizing the Kolmogorov-Smirnov (KS) distance \cite[]{alstott2014powerlaw}, but $x_{\min}$ can also be specified. The KS test is also used to assess the goodness of fit: if the $p$-value of the test is smaller than the conventional significance level of 0.05, one may reject the null hypothesis that a size distribution follows the power law. 

\section{Results} 
\label{sec:results}

\subsection{Waiting time distribution}
\label{S-WT}
\begin{figure} 
\centerline{\hspace*{0.015\textwidth}
\hspace*{-0.03\textwidth}
\includegraphics[width=1.0\textwidth,clip=]{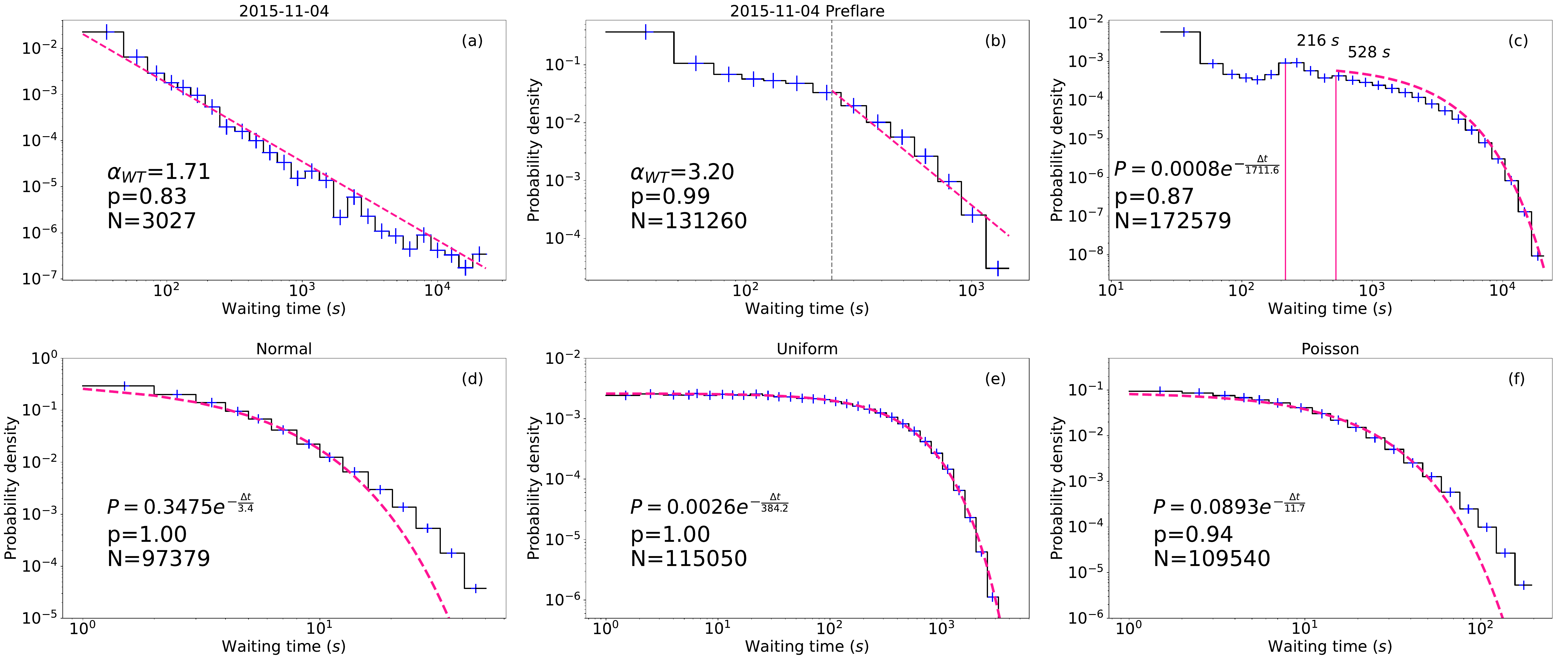}}
\caption{WTDs of an eruptive double-ribbon flare on 2015 November 4. Panel (a) shows the WTD (blue) for the identified flaring pixels, and the MLE fitting with a power-law function (pink). Panel (b) shows the WTD of the same region with the same time interval before the flare. Panel (c) shows the WTD for a quiet-Sun region of the same size and duration as the flaring region. The distribution above the peak at 528~s is fitted by an exponential function with the MLE method. Panel (d-f) show the WTDs from experiments using random numbers generated from uniform, normal, and Poisson distribution functions, respectively. The WTDs are fitted by an exponential function with the MLE method. $p$-values are given by the KS test. $N$ indicates the sample size.} 

\label{F-20151104WTDQS} 
\end{figure}

\begin{figure} 
\centerline{\hspace*{0.015\textwidth}
\hspace*{-0.03\textwidth}
\includegraphics[width=0.95\textwidth,clip=]{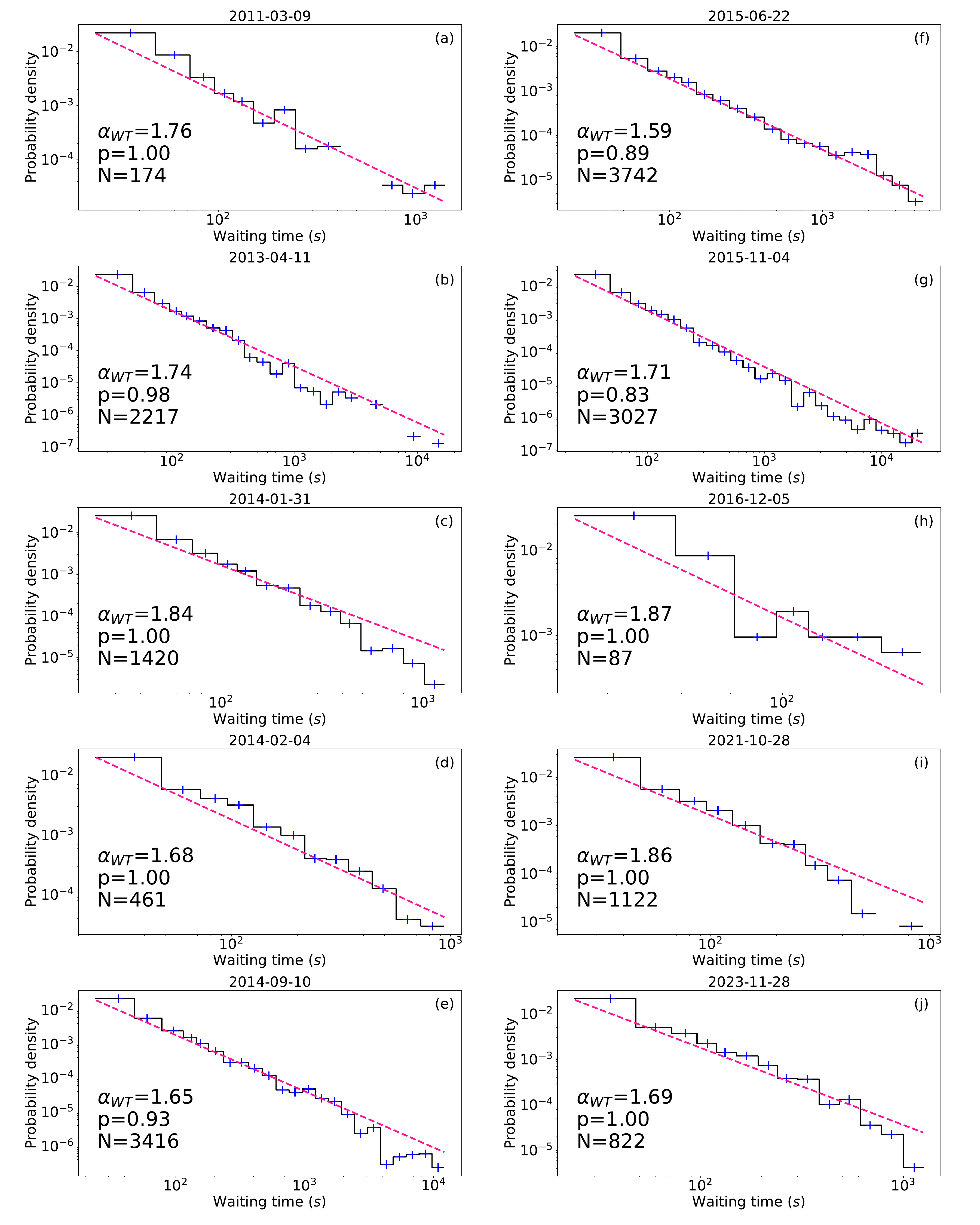}}
\caption{WTD for each of the 10 selected flare events. The distributions (blue) are fitted by a power-law function using the MLE method (red). $p$-values are given by the KS test. $N$ indicates the sample size.}
\label{F-allWTDs} 
\end{figure}

\begin{figure} 
\centerline{\hspace*{0.015\textwidth}
\hspace*{-0.03\textwidth}
\includegraphics[width=\textwidth,clip=]{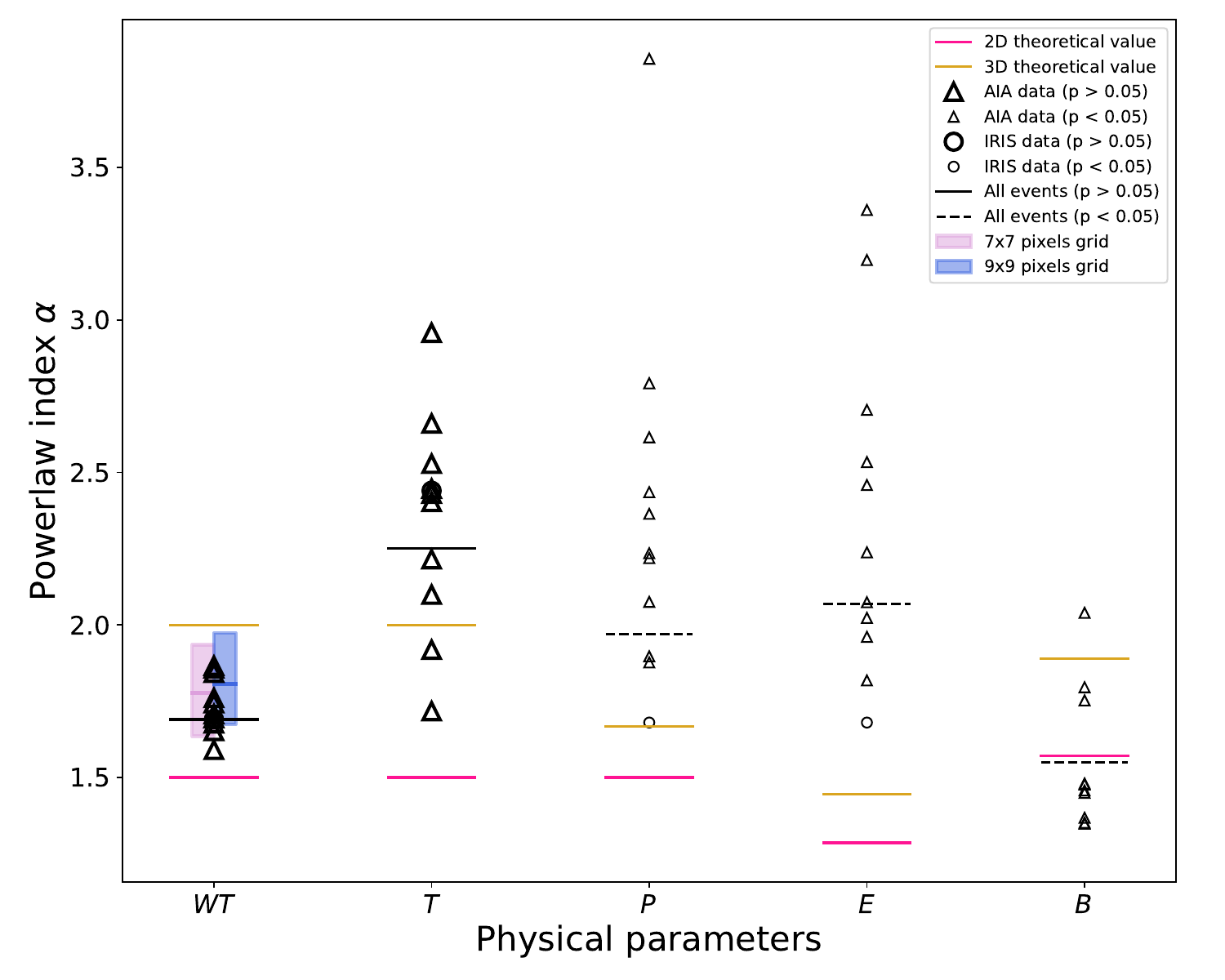}}
\caption{Power-law indices for various physical parameters of flaring pixels in comparison with the SOC model. The triangles (circles) show the indices derived from AIA (IRIS) data (Table~\ref{T-powerlaw}), and the symbol size indicates whether the KS test favors the power-law hypothesis (larger) or not (smaller). Black horizontal line segments show the average indices from all 10 flares (Figure~\ref{F-ALL}), and the line format indicates wether the KS test favors the power-law hypothesis (solid) or not (dashed). The magenta (orange) horizontal line segments mark the predictions of the 2D (3D) standard SOC model (Appendix~\ref{app-theory}), respectively. The purple (blue) vertical bar gives the slope range of WTDs derived from $7\times7$ ($9\times9$) grids (Table~\ref{T-grid}). The horizontal line segment inside the bar indicates the mean value.}
\label{F-slope_vs_theory} 
\end{figure}

The flare ribbons of the M3.7-class flare on 2015 November 4 (Event \#7) are identified by the procedure as explicated in \S\ref{sec:methods}.  The WTD as shown in Fig.~\ref{F-20151104WTDQS}a is obtained by applying a uniform grid of $5\times5$ pixels to the flaring region. It follows a power-law distribution function with a slope of $\sim\,$1.8. The WTDs for the rest of the events also follow power-law distribution functions, as confirmed by KS tests (Fig.~\ref{F-allWTDs}). The power-law indices typically range from 1.5 to 2; the lower and upper bound corresponds to the values predicted by the 2D and 3D standard SOC model (Appendix~\ref{app-theory}), respectively (Fig.~\ref{F-slope_vs_theory}). Fig.~\ref{F-slope_vs_theory} also gives the range of power-law indices of WTDs as obtained from grids of the cell size 7-by-7 pixels (purple) and 9-by-9 pixels (blue). There is a shift toward higher indices as the grid cells become larger. This is because counting waiting times inside a larger cell makes connections between pixels that are located farther away, which yields more short waiting times; consequently the WTD becomes steeper. 

To demonstrate that the power-law-like WTDs do reflect the inherent properties of flare ribbons but not due to the sum of waiting times from individual grid cells, we also obtained WTDs from the same region of the 2015 November 4 event but for the same time duration before the flare (Fig.~\ref{F-20151104WTDQS}b) and from a selected quiet-Sun region of the same size and time duration (Fig.~\ref{F-20151104WTDQS}c). Both WTDs clearly deviate from power-law distributions. The slope of the power-law tail in Fig.~\ref{F-20151104WTDQS}b is over 3, significantly steeper than the theoretical values predicted by the 2D or 3D standard SOC model (Appendix~\ref{app-theory}). The WTD of the quiet-Sun region in Fig.~\ref{F-20151104WTDQS}c shows two peaks at 216 and 528~s, respectively. The larger peak at 216~s might be associated with chromospheric oscillations in the internetwork areas with frequencies exceeding the acoustic cutoff frequency, which is about 5~mHz; the smaller peak at 528~s might be associated with lower-frequency oscillation in the network with periods of 5-20 min \citep{litesDynamicsSolarChromosphere1993}. Above the peak at 528~s, the size distribution can be well fitted by an exponential function, which suggests that brightenings in the quiet Sun may be similar to a Poisson process. 

Further, we carried out experiments by assigning a random peak time to each pixel, so that the distribution of peak times of all the pixels follows a uniform, normal, and Poisson distribution, respectively. The resultant WTDs can also be well fitted by an exponential function in the form of $P=Cf(x)=C\,e^{-\Delta t/\tau}$ (Fig.~\ref{F-20151104WTDQS} (d-f)). Hence, the power-law-like WTDs do not result from any simple linear random process, either.




\subsection{Size distribution of flaring pixel parameters}
\label{S-AIA}
The size distributions of the other four parameters of flaring pixels, including the duration $T$, peak intensity $P$, the area under the light curve $E$, and magnetic field strength $B$, are also calculated, and the corresponding power-law indices are obtained by MLE fittings (Table~\ref{T-powerlaw}). Similar to the WTDs, the size distributions for the duration of flaring pixels generally follows power-law distributions, and the corresponding slopes seem to scatter about the slope predicted by the 3D SOC model (Figure~\ref{F-slope_vs_theory}). In contrast, the size distributions for other parameters generally deviate from power laws, and the estimated slopes are significantly different from those predicted by the standard SOC model. 

\begin{figure} 
\centerline{\hspace*{0.015\textwidth}
\hspace*{-0.03\textwidth}
\includegraphics[width=1.3\textwidth,clip=]{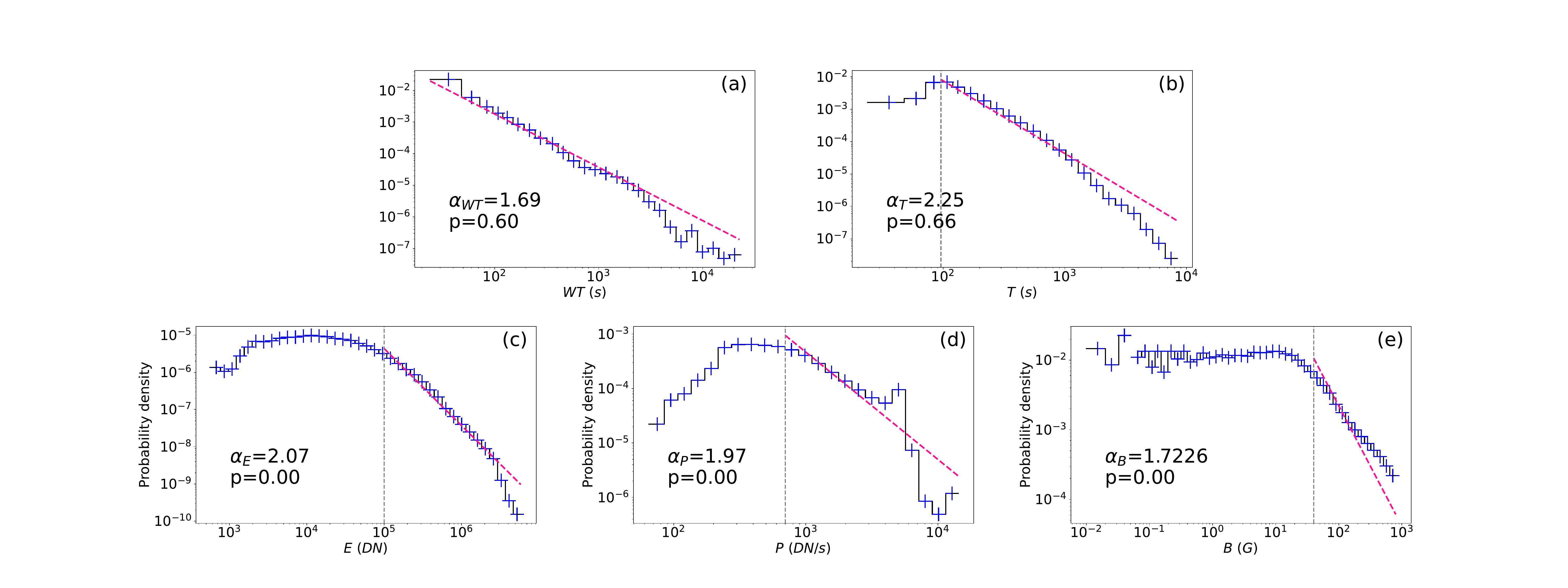}}
\caption{Size distributions for various physical properties of flaring pixels on the ribbons, with synthesized data from all 10 flare events. The size distributions (blue) are fitted by a power-law function using the MLE method (red), with vertical dashed lines marking $x_{\min}$ for power-law fitting unless it coincides with the minimum of the data. $p$-values are given by the KS test.} 
\label{F-ALL} 
\end{figure}

Further, by combining the data from all the ten flare events together, we obtained the `total' size distributions of the physical parameters, and derived for each parameter an `average' slope of all the ten events (Fig.~\ref{F-ALL}). One can see that the size distributions for waiting time and flare duration still follow power laws, but those for other parameters do not, according to the KS tests. Again, the `average' slope for the waiting time falls in between the slopes predicted by the 2D and 3D SOC model, and that for the flare duration is close to what the 3D SOC model predicts (Fig.~\ref{F-slope_vs_theory}). 


\subsection{Size distributions of flare-ribbon parameters in high resolution}

\begin{figure}
\centerline{\hspace*{0.015\textwidth}
\includegraphics[width=0.9\textwidth,clip=]{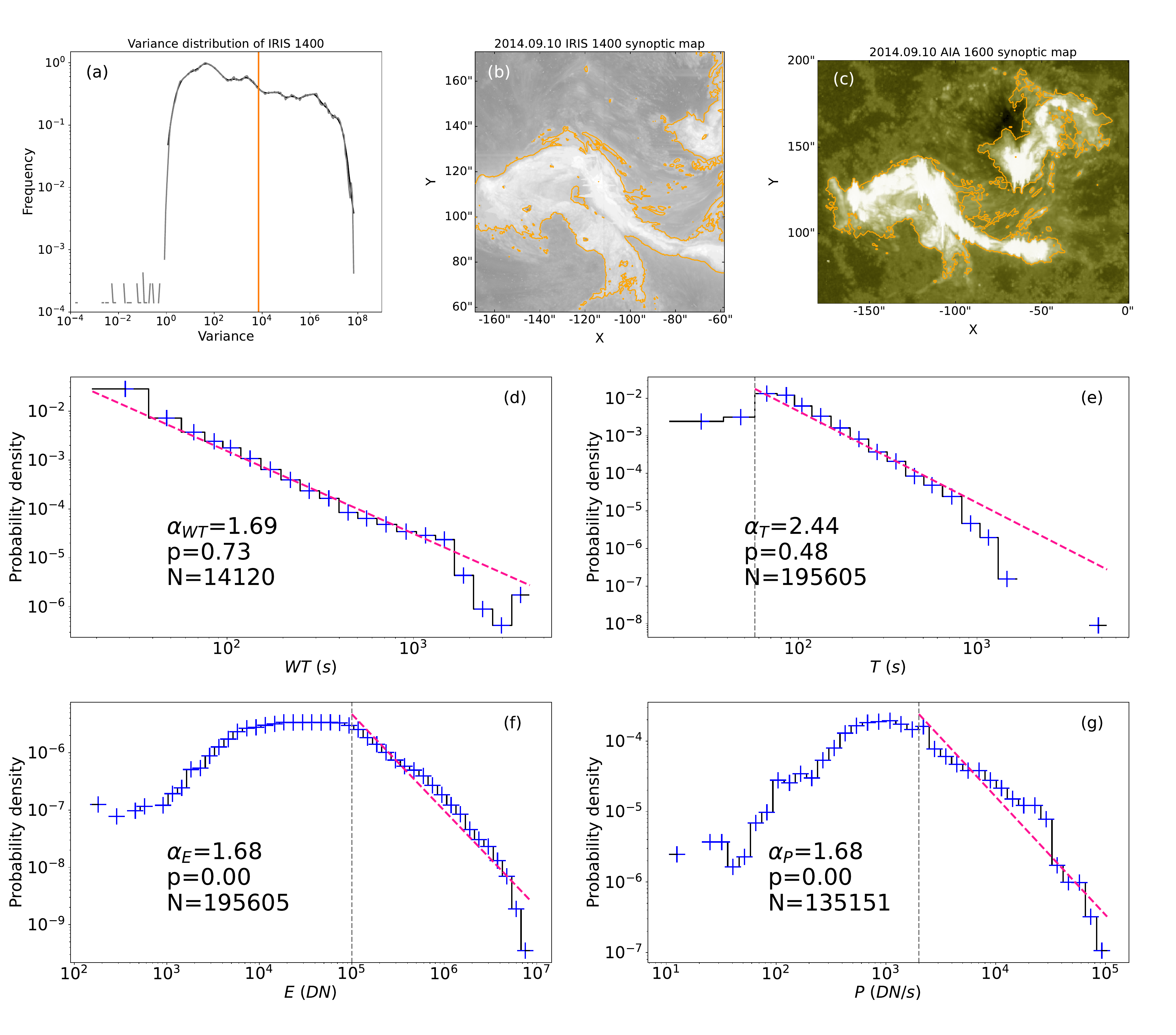}}
\caption{Size distributions for flaring pixels in the 2014 September 10 event. Panels (a - c) show the IRIS 1400~{\AA} flare ribbons (b) as identified by IRIS 1400~{\AA} variance distribution (a) and corresponding flare ribbons in SDO/AIA 1600~{\AA}. The size distributions of $WT$ (d), $T$ (e), $E$ (f), and $P$ (g) are obtained from IRIS data. The size distributions (blue) are fitted by a power-law function using the MLE method (red), with vertical dashed lines marking $x_{\min}$ for power-law fitting unless it coincides with the minimum of the data. $p$-values are given by the KS test. $N$ indicates the sample size. } 
\label{F-IRIS_contour} 
\end{figure}

We applied the approach in Section~\ref{S-WT} to analyze IRIS data of the 2014 September 10 event. The field of view of IRIS covers almost the entire ribbon in the east but only partially the ribbon in the west. Unlike AIA 1600 and 1700~{\AA}, the two available SJI passbands for this period, 1400 and 2796~{\AA}, are quite different in terms of FWHM and temperature responses. Applying the filter ratio of these two passbands does not reduce the background noise. We hence manually chose a threshold from the variance distribution of the IRIS 1400~{\AA} observations (Fig.~\ref{F-IRIS_contour}a) and further apply an intensity threshold so that the extracted fare ribbons visually match those in the synoptic map  (Fig.~\ref{F-IRIS_contour}b). Note that the flare ribbons are better identified with IRIS 1400~{\AA} because it responds less to the flare loops connecting the two ribbons than AIA 1600~{\AA} does (Fig.~\ref{F-IRIS_contour}c). Figure~\ref{F-IRIS_contour} (d--g) show, respectively, the size distributions of the waiting time between flaring pixels (counted within in each individual 5 by 5 pixel cell, \S~\ref{sec:methods}), the duration of the light curves, the energy as approximated by the area below light curves, and the peak intensity of flaring pixels. Similar to the AIA data, the former two distributions are consistent with power laws, but the latter two are not. Further, the power-law slopes obtained with IRIS data are not quite different from those obtained with AIA data (Fig.~\ref{F-slope_vs_theory}). Since the results from AIA and IRIS data are consistent with each other, we tentatively concluded that the obtained size distributions are relatively robust and not sensitive to the spatio-temporal resolution.

\section{Conclusion \& Discussion} \label{sec:conclusion}

In this paper, we detected flare ribbons by combining the variance distribution of the AIA UV filter ratio and the UV intensity threshold. We then investigated the light curve of each flaring pixel on the ribbons to obtain the size distributions of five physical parameters, namely the waiting time, flaring duration, peak intensity, energy, and radial magnetic field strength. Our results showed that the size distributions for waiting time and duration are consistent with power laws, which is a key characteristic of SOC systems, whereas those for other parameters obviously deviate from power laws. These deviations cannot be simply explained as instrumental effects or detection biases, such as weak flare brightenings being missed by the detectors or confused with plages, because waiting time and duration are subject to the same effects but still exhibit clear power-law distributions.

Further, the power-law indices for waiting times seem to be well constrained by the 2D and 3D standard SOC model, and those for the flare duration are more consistent with the 3D SOC model than the 2D model. However, it must be kept in mind that the model assumes that there is no time overlapping between consecutive events but flaring pixels are typically very `busy', and this pileup effect will modify the distribution of both waiting times and durations by steepening the power-law slope by a factor of $\log T_\mathrm{max}/\log\langle \Delta t\rangle$ \cite[]{aschwandenMACROSCOPICDESCRIPTIONGENERALIZED2014}, where $T_\mathrm{max}$ is the upper bound of the power-law range and $\langle \Delta t\rangle$ the mean waiting time. In our cases, the factor is estimated to be in the range of [1.4, 2.0]. 

Alternatively, the observed power-law distributions could be produced by other stochastic processes \cite[e.g.,][]{Aschwanden2021poisson} or MHD turbulence \cite[e.g.,][]{boffettaPowerLawsSolar1999}. For example, the L\'{e}vy flight model of self-similar processes yields a power-law tail $\sim\Delta t^{-(1+\mu)}$, with $0<\mu\le2$ \cite[]{lepretiSolarFlareWaiting2001}, predicting a power-law slope in the wide range of [1, 3]. A nonstationary Poisson process exhibits a power-law distribution with the slope in a similar range of $1\lesssim\alpha\lesssim3$ \cite[]{Wheatland&Litvinenko2002,aschwandenRECONCILIATIONWAITINGTIME2010}. \cite{Aschwanden2021poisson} further narrowed down this range to [2, 2.5] in the nonlinear regime, which compares favorably to the observed exponents (Fig.~\ref{F-slope_vs_theory}). Thus, without excluding the L\'{e}vy process or turbulence, we tentatively argue that the observed WTDs favors a nonstationary Poisson or an avalanche-like process in the flaring current sheet in a temporal sense, but in other aspects, i.e., space- and energy-wise, this process is likely modulated by other physical processes or the structures of the flaring current sheet.  

We obtained the WTDs by partitioning time-elapsed images with uniform grids and collecting waiting times from each individual grid cells. Consequently different WTDs are derived from different regions and differently time periods. The WTD in the flaring region as characterized by power-laws is distinct from that in the same region before the flare, which deviates from power laws and falls off more rapidly, or that in the quiet-Sun region, which is close to an exponential distribution and characterized by two peaks at about 216~s and 528~s, respectively. The larger (smaller) peak at 216~s (528~s) might be associated with chromospheric oscillations in the internetwork (network) areas. Hence, the WTDs are shaped by both the temporal randomness and the spatial distribution and correlation of brightening events, under the assumption that there are no long-range spatio-temporal correlations between them. 

\begin{acknowledgments}
This work was supported by the National Key R\&D Program of China (2022YFF0503002), the Strategic Priority Program of the Chinese Academy of Sciences (XDB0560102), and the NSFC (42274204, 12373064, 42188101, and 11925302). 
\end{acknowledgments}

\begin{appendices}
\section{Power-law indices predicted by the standard SOC model} \label{app-theory}
The standard SOC model, which is the statistical fractal-diffusive avalanche model of a slowly driven SOC system \citep[FD-SOC;][]{Aschwanden2012A&A,aschwandenMACROSCOPICDESCRIPTIONGENERALIZED2014,Aschwanden2022ApJ}, is based on a few assumptions. First, the universal size distribution $N(L)$ of length scales $L$ is given by the scale-free probability conjecture, i.e.,  
\[N(L)\,dL \propto L^{-d}dL,\]
where $d$ is the Euclidean space dimension. Second, the system has a fractal dimension $D_d$, whose upper limit is the Euclidean dimension $d$. $D_d$, which ranges between $d-1$ and $d$, can be approximated as follows,
\[D_d \simeq \frac{\min(D_d)+\max(D_d)}{2}=d-\frac12.\]
Third, the SOC system exhibits a diffusive behavior such that the time duration $T$ of an avalanche is linked to the length scale,
\[ T\propto L^{2/\beta},\]
where $\beta =1 $ for classical diffusion. Fourth, a proportional relationship exists between the observed mean flux $F$ and the emitting surface or volume, such that
\[F \propto L^{\gamma D_d},\] 
which is generally valid for incoherent processes. Often $\gamma$ is set to unity for simplicity. The peak flux $P$ is achieved when $D_d$ is replaced by $d$ for the flux $F$, i.e., 
\[P \propto L^{\gamma d}.\]

Following \cite{Aschwanden2022ApJ},  we obtain the expected power-law indices for the size distributions of spatial length $L$, fractal area $A_f$, fractal volume $V_f$, duration $T$, waiting time $WT$ (assumed to have the same size distribution as $T$),  peak flux $P$, energy $E=F\times T$ (approximated by the area under the light curves of observed intensity or flux), and magnetic field $B$ (with $B^2\propto E$):
\begin{equation*}
    \begin{split}
        &{\alpha}_{WT}={\alpha}_T=1+\frac{(d-1){\beta}}{2};\\
        &{\alpha}_P=1+\frac{d-1}{{\gamma}d};\\
        &{\alpha}_E=1+\frac{d-1}{{\gamma}D_d+2/{\beta}};\\
        &{\alpha}_B=1+\frac{2(d-1)}{{\gamma}D_d+2/\beta}.
    \end{split}
\end{equation*}
The values for the 2D ($d=2$) and 3D ($d=3$) occasions are given below, with $\beta=1$ and $\gamma=1$:
\begin{center}
\tabcolsep=0.3cm
\begin{tabular}{c|cccc}
 &$\alpha_{WT}$ ($\alpha_{T}$) & $\alpha_P$ & $\alpha_E$ & $\alpha_B$  \\ \hline
$d=2$ & $\frac32$ & $\frac32$ & $\frac97$ & $\frac{11}{7}$ \\
$d=3$ & 2 & $\frac53$ & $\frac{13}{9}$ & $\frac{17}{9}$
\end{tabular}
\end{center}

\end{appendices}


\bibliography{final}
\bibliographystyle{aasjournal}

\begin{deluxetable}{lcccccccc} 
\tablecaption{List of flare events and properties of flare ribbons} \label{T-list_of_events}
\tablecolumns{9}	
\tablehead{\multirow{2}*{Event} & \multirow{2}*{Time} & \multirow{2}*{Location} & 
    \multirow{2}*{NOAA} & \multirow{2}*{Class} & \multirow{2}*{E/C} & \colhead{$A$} & \colhead{$\Phi$} & \colhead{$\overline{B}_r$} \\
    \colhead{} & \colhead{} & \colhead{} & \colhead{} & \colhead{} & \colhead{} & \colhead{(Mm$^2$)} & \colhead{($10^{20}$Mx)} & \colhead{(Gauss)} 
    }         
\startdata                
 1     &  2011-03-09T10:35      &  N08W11  &  11166  &  M1.7 &  C  & $289.35^{+44.88}_{-81.05}$          & $7.40^{+1.12}_{-2.07}$        & $511.65^{-1.53}_{+0.77}$        \\
 2     &  2013-04-11T06:55      &  N07E13  &  11719  &  M6.5 &  E  & $2960.17^{+461.49}_{-1185.81}$      & $22.24^{+3.34}_{-8.04}$       & $150.23^{-0.77}_{+9.82}$        \\
 3     &  2014-01-31T15:32      &  N07E34  &  11968  &  M1.1 &  E  & $1887.79.70^{+55.86}_{-293.71}$       & $12.29^{+0.48}_{-1.95}$       & $130.20^{+1.17}_{-0.53}$        \\
 4     &  2014-02-04T03:57      &  S14W07  &  11967  &  M5.2 &  C  & $335.94^{+25.19}_{-53.02}$          & $19.07^{+1.44}_{-2.86}$       & $1135.48^{+0.53}_{+10.64}$        \\
 5     &  2014-09-10T17:21      &  N11E05  &  12158  &  X1.6 &  E  & $3145.75^{+46.21}_{-142.40}$       & $67.62^{+0.77}_{-1.87}$       & $429.90^{-1.38}_{+7.95}$        \\
 6     &  2015-06-22T17:39      &  N13W06  &  12371  &  M6.5 &  E  & $3609.88^{+318.70}_{-962.74}$     & $69.42^{+3.06}_{-10.24}$     & $384.61^{-15.64}_{+62.48}$        \\
 7     &  2015-11-04T13:31      &  N08W02  &  12443  &  M3.7 &  E  & $3465.02^{+62.68}_{-111.16}$       & $32.19^{+0.58}_{-1.12}$       & $185.82^{-0.02}_{-0.49}$        \\
 8     &  2016-12-05T05:57      &  S08W21  &  12615  &  C1.2 &  E  & $151.87^{+0.19}_{-0.00}$          & $1.16^{+0.00}_{-0.00}$        & $153.14^{-0.16}_{+0.00}$        \\
 9     &  2021-10-28T15:17      &  S28W01  &  12887  &  X1.0 &  E  & $1360.03^{+130.85}_{-135.21}$       & $20.78^{+1.61}_{-1.76}$       & $305.59^{-5.17}_{+4.96}$        \\
 10    &  2023-11-28T19:35      &  S16W00  &  13500  &  M9.8 &  E  & $749.32^{+129.34}_{-158.88}$        & $23.63^{+2.70}_{-3.90}$       & $630.78^{-31.34}_{+37.61}$        \\    
\enddata
\tablecomments{`E' and `C' represent eruptive and confined flares, respectively. $A$ indicates the ribbon area, $\Phi$ indicates the unsigned magnetic flux, which is an average of positive flux $\Phi_p$ and absolute values of negative flux $\Phi_n$, i.e., $\dfrac12(\Phi_p + |\Phi_n|)$, through flare ribbons, and $\overline{B}_r$ is the mean strength of the radial component of photospheric magnetic fields over flare ribbons. For the flare-ribbon properties, the superscript (subscript)  indicates the difference between the values obtained from the lower (upper) threshold and those from the median threshold.} 
\end{deluxetable}

\begin{deluxetable}{c c c c c c c }
\centering
\tablecaption{Power-law indices for the size distributions of flare-ribbon parameters of the 10 selected events.} \label{T-powerlaw}
\tablecolumns{7}
\tablehead{ & Event  & $\alpha_{WT}$ (p-value)  & $\alpha_T$ (p-value)  & $\alpha_P$ (p-value) & $\alpha_{E} $ (p-value)  & $\alpha_{B}$ (p-value)}

\startdata   
1  & 2011.03.09 & $1.76^{+0.02}_{-0.01}$ (1.00) & $2.53^{-0.05}_{+0.15}$ (0.99) & $2.79^{+0.08}_{-0.12}$ (0.00) & $2.24^{+0.04}_{-0.05}$ (0.00) & $1.45^{-0.01}_{-0.00}$ (0.00) \\
2  & 2013.04.11 & $1.74^{-0.03}_{+0.02}$ (0.98) & $2.43^{-0.11}_{+0.12}$ (0.88) & $2.24^{+0.05}_{-0.24}$ (0.00) & $2.02^{+0.03}_{-0.08}$ (0.00) & $1.80^{-0.01}_{-0.02}$ (0.00) \\
3  & 2014.01.31 & $1.84^{-0.00}_{+0.01}$ (1.00) & $2.66^{+0.00}_{+0.01}$ (0.95) & $2.08^{+0.01}_{-0.06}$ (0.00) & $2.46^{+0.01}_{-0.03}$ (0.00) & $2.04^{-0.01}_{+0.00}$ (0.00) \\
4  & 2014.02.04 & $1.68^{-0.00}_{+0.01}$ (1.00) & $1.72^{-0.00}_{+0.00}$ (0.11) & $1.90^{+0.02}_{-0.03}$ (0.00) & $1.96^{+0.02}_{-0.04}$ (0.01) & $1.35^{-0.01}_{-0.01}$ (0.00) \\
5  & 2014.09.10 & $1.65^{-0.00}_{+0.00}$ (0.93) & $2.10^{-0.00}_{+0.00}$ (0.61) & $2.36^{+0.01}_{-0.02}$ (0.00) & $1.82^{+0.00}_{-0.01}$ (0.00) & $1.48^{+0.03}_{-0.02}$ (0.00) \\
 & IRIS &$1.69$ (0.73) &$2.44$ (0.48) &$1.68$ (0.00) &$1.68$ (0.00) \\
6  & 2015.06.22 & $1.59^{+0.01}_{-0.00}$ (0.89) & $2.22^{+0.01}_{-0.04}$ (0.94) & $1.88^{+0.03}_{-0.11}$ (0.00) & $2.07^{+0.01}_{-0.06}$ (0.00) & $1.37^{+0.02}_{-0.02}$ (0.00) \\
7  & 2015.11.04 & $1.71^{-0.00}_{+0.00}$ (0.83) & $2.40^{-0.00}_{+0.00}$ (0.99) & $2.62^{+0.01}_{-0.02}$ (0.00) & $2.53^{+0.00}_{-0.00}$ (0.00) & $1.46^{-0.03}_{+0.02}$ (0.00) \\
8  & 2016.12.05 & $1.87^{+0.00}_{+0.00}$ (1.00) & $2.96^{-0.00}_{-0.00}$ (1.00) & $3.86^{+0.00}_{-0.00}$ (0.01) & $3.20^{+0.00}_{-0.00}$ (0.00) & $1.75^{+0.03}_{+0.04}$ (0.00) \\
9  & 2021.10.28 & $1.86^{-0.00}_{+0.00}$ (1.00) & $2.45^{-0.03}_{+0.05}$ (1.00) & $2.44^{+0.02}_{-0.02}$ (0.00) & $2.71^{+0.03}_{-0.03}$ (0.00) & $1.48^{+0.00}_{-0.00}$ (0.00) \\
10 & 2023.11.28 & $1.69^{-0.00}_{+0.03}$ (1.00) & $1.92^{-0.02}_{+0.05}$ (0.41) & $2.22^{+0.04}_{-0.07}$ (0.00) & $3.36^{+0.03}_{-0.05}$ (0.00) & $1.35^{+0.01}_{-0.01}$ (0.00) 
\enddata
\tablecomments{The power-law indices are given for the flare ribbons identified by the median threshold (\S~\ref{sec:methods}). The superscript (subscript) indicates the difference between the power-law index for the lower (upper) and that for the median threshold. $p$-values from the K-S test for the median threshold are given in the brackets. }
\end{deluxetable}

\begin{deluxetable}{ccccc}
\centering
\tablecaption{Power-law indices of WTDs with different grid patterns} 
\label{T-grid}
\tablecolumns{5}
\tablehead{ & Event & $5\times5$ & $7\times7$ & $9\times9$}         
\startdata
1  & 2011.03.09 & $1.76^{+0.02}_{-0.01}$ (1.00) & $1.78^{+0.04}_{-0.02}$ (0.99) & $1.77^{+0.05}_{-0.02}$ (1.00) \\
2  & 2013.04.11 & $1.74^{-0.03}_{+0.02}$ (0.98) & $1.79^{-0.03}_{+0.01}$ (0.85) & $1.82^{-0.04}_{+0.01}$ (0.65) \\
3  & 2014.01.31 & $1.84^{-0.00}_{+0.01}$ (1.00) & $1.87^{-0.00}_{+0.02}$ (0.99) & $1.91^{+0.00}_{+0.01}$ (0.97) \\
4  & 2014.02.04 & $1.68^{-0.00}_{+0.01}$ (1.00) & $1.72^{+0.01}_{+0.02}$ (1.00) & $1.78^{+0.00}_{-0.00}$ (1.00) \\
5  & 2014.09.10 & $1.65^{-0.00}_{+0.00}$ (0.93) & $1.69^{-0.00}_{+0.00}$ (0.67) & $1.73^{-0.00}_{+0.00}$ (0.55) \\
6  & 2015.06.22 & $1.59^{+0.01}_{-0.00}$ (0.89) & $1.64^{+0.01}_{-0.01}$ (0.75) & $1.67^{+0.01}_{-0.00}$ (0.50) \\
7  & 2015.11.04 & $1.71^{-0.00}_{+0.00}$ (0.83) & $1.74^{-0.00}_{+0.00}$ (0.66) & $1.74^{-0.01}_{+0.01}$ (0.31) \\
8  & 2016.12.05 & $1.87^{+0.00}_{+0.00}$ (1.00) & $1.94^{+0.00}_{+0.00}$ (1.00) & $1.97^{+0.00}_{+0.00}$ (1.00) \\
9  & 2021.10.28 & $1.86^{-0.00}_{+0.00}$ (1.00) & $1.91^{-0.01}_{-0.01}$ (1.00) & $1.91^{-0.00}_{+0.01}$ (1.00) \\
10 & 2023.11.28 & $1.69^{-0.00}_{+0.03}$ (1.00) & $1.71^{-0.00}_{+0.02}$ (1.00) & $1.76^{-0.01}_{+0.01}$ (1.00) 
\enddata
\end{deluxetable}




\end{document}